\documentclass[aps,prl,twocolumn,showpacs,preprintnumbers,amsmath,amssymb]{revtex4}
\usepackage{graphicx}
\usepackage{dcolumn}
\usepackage{bm}

\newcommand{\tc}{{,~}}
\begin{document}
\title{$Q^2$-Dependence of the Neutron Spin Structure Function $g_2^n$ at Low $Q^2$}

\author{
K.~Kramer,$^{20}$
D.~S.~Armstrong,$^{20}$
T.~D.~Averett,$^{18,20}$
W.~Bertozzi,$^{12}$
S.~Binet,$^{19}$
C.~Butuceanu,$^{20}$
A.~Camsonne,$^2$
G.~D.~Cates,$^{19}$
J.-P.~Chen,$^{18}$
Seonho~Choi,$^{17}$
E.~Chudakov,$^{18}$
F.~Cusanno,$^6$
A.~Deur,$^{19}$
P.~Djawotho,$^{20,\dagger}$
D.~Dutta,$^{12}$
J.~M.~Finn,$^{20}$
H.~Gao,$^{12}$
F.~Garibaldi,$^{6}$
O.~Gayou,$^{12}$
R.~Gilman,$^{18,15}$
A.~Glamazdin,$^{9}$
V.~Gorbenko,$^{9}$
K.~A.~Griffioen,$^{20}$
J.-O.~Hansen,$^{18}$
D.~W.~Higinbotham,$^{12}$
W.~Hinton,$^{14}$
T.~Horn,$^{10}$
C.W.~de~Jager,$^{18}$
X.~Jiang,$^{15}$
W.~Korsch,$^8$
J.~LeRose,$^{18}$
D.~Lhuillier,$^{16}$
N.~Liyanage,$^{18}$
D.~J.~Margaziotis,$^1$
K.~McCormick,$^7$
Z.-E.~Meziani,$^{17}$
R.~Michaels,$^{18}$
B.~Milbrath,$^3$
B.~Moffit,$^{20}$
S.~Nanda,$^{18}$
C.~F.~Perdrisat,$^{20}$
R.~Pomatsalyuk,$^9$
V.~Punjabi,$^{13}$
B.~Reitz,$^{18}$
J.~Roche,$^{20}$
R.~Roch\'{e},$^4$
M.~Roedelbronn,$^5$
N.~Savvinov,$^10$
J.~Secrest,$^{20}$
J.~Singh,$^{19}$
S.~\v{S}irca,$^{12}$
K.~Slifer,$^{17}$
P.~Solvignon,$^{17}$
D.~J.~Steiner,$^{20}$
R.~Suleiman,$^{12}$
V.~Sulkosky,$^{20}$
A.~Tobias,$^{19}$
A.~Vacheret,$^{11}$
Y.~Xiao,$^{12}$
X.~Zheng,$^{12}$
J.~Zhou,$^{12}$
L.~Zhu,$^{12}$
X.~Zhu,$^{20}$
P.~A.~\.Zo{\l}nierczuk,$^8$
}
\affiliation{
\baselineskip 2 pt
\vskip 0.3 cm
{\small{
{$^1$California State University\tc Los Angeles\tc Los Angeles\tc California
90032\tc USA} \break
{$^2$Universit\'{e} Blaise Pascal Clermont-Ferrand et CNRS/IN2P3 LPC
63\tc 177 Aubi\`{e}re Cedex, France} \break
{$^3$Eastern Kentucky University\tc Richmond, Kentucky 40475\tc  USA} \break
{$^4$Florida State University\tc Tallahassee\tc Florida 32306\tc  USA} \break
{$^5$University of Illinois\tc Urbana\tc Illinois 61801\tc  USA} \break
{$^6$Istituto Nazionale di Fisica Nucleare\tc Sezione Sanit\`a\tc
00161 Roma\tc Italy} \break
{$^7$Kent State University\tc Kent\tc Ohio 44242\tc  USA} \break
{$^8$University of Kentucky\tc Lexington\tc Kentucky 40506\tc  USA} \break
{$^9$Kharkov Institute of Physics and Technology\tc Kharkov 61108\tc Ukraine} \break
{$^{10}$University of Maryland\tc College Park\tc Maryland 20742\tc  USA} \break
{$^{11}$University of Massachusetts Amherst\tc Amherst\tc Massachusetts 01003\tc  USA}
\break
{$^{12}$Massachusetts Institute of Technology\tc Cambridge\tc Massachusetts 02139\tc  USA}
\break
{$^{13}$Norfolk State University\tc Norfolk\tc Virginia 23504\tc  USA} \break
{$^{14}$Old Dominion University\tc Norfolk\tc Virginia 23529\tc  USA} \break
{$^{15}$Rutgers\tc The State University of New Jersey\tc Piscataway\tc
New Jersey 08855\tc  USA} \break
{$^{16}$CEA Saclay\tc DAPNIA/SPhN\tc F-91191 Gif sur Yvette\tc France}
\break
{$^{17}$Temple University\tc Philadelphia\tc Pennsylvania 19122\tc  USA} \break
{$^{18}$Thomas Jefferson National Accelerator Facility\tc Newport
News\tc Virginia 23606\tc  USA} \break
{$^{19}$University of Virginia\tc Charlottesville\tc Virginia 22904\tc  USA} \break
{$^{20}$College of William and Mary\tc Williamsburg\tc Virginia 23187\tc  USA} \break
}}
}

\date{\today}
\begin{abstract}
We present the first measurement of the $Q^2$-dependence of the neutron spin structure function $g_2^n$ at five kinematic points covering $0.57$ $({\rm{GeV}}/c)^2$ $\leq Q^2 \leq 1.34$  $({\rm{GeV}}/c)^2$ at $x\simeq 0.2$.  Though the naive quark-parton model predicts $g_2=0$, non-zero values occur in more realistic models of the nucleon which include quark-gluon correlations, finite quark masses or orbital angular momentum.  When scattering from a non-interacting quark,  $g_2^n$ can be predicted using next-to-leading order fits to world data for $g_1^n$.  Deviations from this prediction provide an opportunity to examine QCD dynamics in nucleon structure.   Our results show a positive deviation from this prediction at lower $Q^2$, indicating that contributions such as quark-gluon interactions may be important.  Precision data obtained for $g_1^n$  are consistent with next-to-leading order fits to world data. 
\end{abstract}
\maketitle

Over the past 30 years, significant progress has been made in understanding the spin structure of the nucleon through measurements using polarized deep-inelastic lepton scattering (DIS).  Most of these experiments were focused on precise measurements of the spin structure function $g_1$.  In the naive quark-parton model (QPM), $g_1$ is directly related to the contributions of the individual quark flavors to the overall spin of the nucleon (see e.g. Ref.~\cite{Thomas}).  Sum rules based on this simple model have provided fertile ground for understanding the origin of the nucleon spin in terms of quark degrees of freedom.  In addition, next-to-leading-order (NLO)  analyses of the world $g_1$ data (see e.g. Refs.~\cite{BB, AAC03}) have provided indirect information about the role of  gluons in the nucleon's spin.

The QPM is expected to be valid in the scaling limit, where the four-momentum transfer squared $-Q^2$ and energy transferred $\nu$ approach infinity.  As $Q^2$ becomes large, the electron-nucleon interaction can be described by scattering from a massless, non-interacting quark carrying a fraction $x=Q^2/2M\nu$ of the nucleon's momentum ($M$ is the nucleon mass).   At finite $Q^2$ and $\nu$, effects such as gluon Bremsstrahlung, vacuum polarization and vertex corrections can be accurately calculated using perturbative QCD.  For $g_1$, non-perturbative QCD processes, such as quark-gluon and quark-quark correlations, are suppressed relative to the asymptotically-free contributions by factors of $1/Q$ and $1/Q^2$, respectively.

Polarized DIS also provides information about a second spin structure function, $g_2$, which is identically zero in the QPM~\cite{Manohar}.  Interest in $g_2$ arises because, unlike $g_1$, contributions from certain non-perturbative processes enter at the same order in $Q^2$ as the asymptotically-free contributions.  An appropriate formalism for understanding $g_2$ is the operator product expansion (OPE)~\cite{OPE1,OPE2} which is a model-independent approach based directly on QCD.   Here, the unknown hadronic currents relevant for polarized DIS are expanded in terms of quark and gluon operators and grouped by factors of $(1/Q)^{\tau-2}$, where $\tau=2,3,4,...$ is known as the {\em{twist}} of the operator.   Both $g_1$ and $g_2$ contain a contribution from a twist-2 operator that corresponds to scattering from a massless,  non-interacting quark.   Operators with higher-twist represent contributions from non-perturbative processes such as quark-quark and quark-gluon correlations, and from quark mass effects.  Because these correlations are responsible for quark confinement, higher-twist effects must be included in any realistic model of the nucleon~\cite{OPE2}.  

Using the fact that $g_1$ and $g_2$ contain the same twist-2 operator, Wandzura and Wilczek~\cite{g2ww} derived the following expression for the asymptotically-free contribution to $g_2$, in terms of $g_1$,
\begin{equation}
g_2^{WW}(x,Q^2)=-g_1(x,Q^2)+\int_x^1\frac{g_1(x,Q^2)}{x}dx.
\label{eq:g2ww}
\end{equation}
The world data for $g_1$ cover a broad range in $x$ and $Q^2$ with relatively high precision.  Polarized parton distributions can be extracted from NLO fits to the data and evolved to different values of $Q^2$ using the Dokshitzer-Gribov-Lipatov-Alarelli-Paresi (DGLAP) procedure~\cite{DGLAP1, DGLAP2, DGLAP3, DGLAP4}.  Because the world data are at relatively large $Q^2$, where higher-twist effects should be negligible, these evolved parton distributions allow one to calculate the twist-2 part of $g_1$, and therefore $g_2^{WW}$, in most kinematic regions accessible today.   Precise measurements of $g_2$ at specific values of $x$ and $Q^2$ can be compared to $g_2^{WW}$, providing a unique opportunity to cleanly isolate higher-twist contributions. 

Previous measurements of  $g_2$~\cite{E143, E154, E155,E155x} were aimed at testing OPE sum rules and covered a wide range in $x$, at an average $Q^2$ of 5 $({\rm{GeV}}/c)^2$.   Proton data show general agreement with $g_2^{WW}$ but lack the precision needed to make a definitive statement about the size of higher-twist effects by direct comparison.   Neutron results have much larger uncertainties and cannot distinguish between $g_2^{WW}$ and $g_2=0$.  We present a new measurement of the $Q^2$-dependence of $g_2$ for the neutron at low $Q^2$, while keeping $x$ approximately constant.  With statistical uncertainties more than an order of magnitude smaller than existing data, these results allow, for the first time, a precise comparison with $g_2^{WW}$ to study the $Q^2$-dependence of higher-twist effects.

Longitudinally-polarized electrons were scattered from a polarized $^3$He target in Hall A~\cite{HALLA} at the Thomas Jefferson National Accelerator Facility (Jefferson Lab) through the inclusive process $\vec{^3{\rm{He}}}(\vec{e},e')$.  We measured $g_2$  at five values of $Q^2$ between $0.57$ $({\rm{GeV}}/c)^2$ and $1.34$ $({\rm{GeV}}/c)^2$, with $x\simeq0.2$.  The kinematics for this $x$ are well-matched to the beam energies available at Jefferson Lab and are in a region where $g_2^{WW}$ is relatively large.  Low  $Q^2$ provides access to the region where higher-twist effects are expected to become important.  The invariant mass squared of the photon-nucleon system, $W^2=M^2 + 2M\nu-Q^2$, was kept $>3.8$ $({\rm{GeV}}/c)^2$ to minimize resonance contributions.  The spin of the $^3$He nuclei could be oriented either longitudinal or transverse to the incoming electron spin.  

The parallel and perpendicular cross-section differences, $\Delta\sigma_{\parallel}$ and $\Delta\sigma_{\perp}$, are defined in terms of helicity-dependent cross-sections $\sigma^{ij}$ as,
\begin{equation}
\Delta\sigma_{\parallel}=\sigma^{\downarrow \Uparrow}-\sigma^{\uparrow \Uparrow},\;\;\;\;
\Delta\sigma_{\perp}=\sigma^{\downarrow \Rightarrow}-\sigma^{\uparrow \Rightarrow},
\end{equation}
where the single and double arrows refer to the beam and target spin directions, respectively, relative to the incident electron momentum direction.  An up (down) arrow refers to spin aligned parallel (anti-parallel) to the momentum.  The double right arrow refers to target polarization in the scattering plane, perpendicular to the incident electron momentum, with the scattered electron detected on the side of the beam line towards which the target spin points.  These cross-section differences are related to $g_1$ and $g_2$ by~\cite{Thomas},
\begin{eqnarray}
\Delta\sigma_{\parallel}&=&\frac{4\alpha^2E'}{Q^2EM\nu}\big[(E+E'\cos \theta) g_1(x,Q^2)\nonumber\\
&-&2xMg_2(x,Q^2)\big],\label{eq:xsec_par}\\
\nonumber\\
\Delta\sigma_{\perp} &=&\frac{4\alpha^2E'}{Q^2EM\nu}E'\sin \theta\big[g_1(x,Q^2)\nonumber\\
&+&\frac{4xEM}{Q^2}g_2(x,Q^2)\big],
\label{eq:xsec_perp}
\end{eqnarray}
where $E$ and $E'$ are the incident and scattered electron energies and  $\theta$ is the laboratory scattering angle.

Polarized electrons were produced by photo-emission from a strained GaAs crystal using circularly polarized laser light.  The average beam polarization was $P_b=(76\pm3)\%$ as measured using both M{\o}ller and Compton polarimeters~\cite{HALLA} and the average beam current was $12.0$ $\mu$A, with an uncertainty of $\pm1\%$.  The helicity of the electron beam was flipped on a pseudo-random basis at a rate of 30 Hz to minimize helicity-correlated systematic effects.  The beam energy ranged from $3.5$ GeV to $5.7$ GeV and was measured with a relative uncertainty below $10^{-3}$ using both elastic electron-proton scattering and a magnetic field measurement.

Scattered particles were detected by either one of a pair of magnetic spectrometers arranged symmetrically on either side of the beam line~\cite{HALLA}.  The momentum of the particles was determined by reconstructing their trajectories using drift chambers.  Electrons were identified using a threshold gas Cherenkov detector and a two-layer lead-glass electromagnetic calorimeter.   The largest background was from $\pi^-$ photo-production in the target.  The ratio of pion rate to electron rate detected in the spectrometer was $<3.3$.  The ratio of the pion asymmetry to the electron asymmetry was $<4.1$.  The pion rejection factor was $>10^4:1$ with an electron detection efficiency of $\geq97\%$, which was sufficient to reduce the pion contamination to a negligible level.   Electrons scattered in the entrance and exit windows of the target cell were removed using software cuts. 

To study polarized neutrons,  a target containing $^3$He nuclei was polarized using spin-exchange optical pumping~\cite{SEOP}.    The  $^3$He ground state is dominated by the $S$-state in which the two proton spins are anti-aligned and the spin of the nucleus is carried entirely by the neutron.  
 The $^3$He is contained in a sealed, two-chambered, aluminosilicate glass cell, along with a small quantity of N$_2$ and Rb to aid in the polarization process.   Polarized $^3$He is produced in the spherical upper chamber by first polarizing Rb atoms with optical pumping.  These atoms can transfer their spin to the $^3$He nucleus during binary collisions.   Incident electrons scatter from the polarized $^3$He in the cylindrical lower chamber which is 40 cm long with side walls of thickness $\approx 1.0$ mm and end windows of thickness  $\approx 130$  $\mu$m.  The $^3$He density as seen by the  beam is $2.9\times 10^{20}$/cm$^3$.  The average in-beam target polarization was $P_t=(40.0\pm1.4)\%$  as measured using both nuclear magnetic resonance~\cite{NMR} and electron paramagnetic resonance~\cite{EPR}.

The $^3$He spin structure functions $g_1$ and $g_2$ were obtained using our measured cross-section differences and Eqs.~(\ref{eq:xsec_par}) and~(\ref{eq:xsec_perp}).  The ratios of these cross-section differences to the total unpolarized cross-section are defined as the longitudinal and transverse asymmetries, respectively.  To achieve our desired statistical precision on $g_2^n$,   the raw (uncorrected) transverse asymmetries  were measured to a statistical uncertainty of $\sim10^{-4}$.   It was also necessary to keep sources of false asymmetries at, or below, this level.   A feedback system was used to keep helicity-dependent beam charge asymmetry below $50\times 10^{-6}$ for a typical run.   Quasi-elastic polarized electron scattering from an unpolarized $^{12}$C target was used to measure the false asymmetry and yielded $(67\pm46)\times 10^{-6}$.  Misalignment of the target polarization direction was typically less than $\pm0.3^{\circ}$, implying a negligible contribution from $\Delta\sigma_{\parallel}$ when measuring $\Delta\sigma_{\perp}$.  For each kinematic setting, equal quantities of data were collected with the overall sign of the electron helicities reversed and/or the target polarization rotated by $180^{\circ}$.  The absolute value of the measured asymmetries for each of the beam helicity and target spin combinations were consistent with each other within the statistical uncertainties, indicating that there were no significant false asymmetries.

The raw cross-section differences were determined with a relative uncertainty of $6\%$ and were checked against elastic data taken during this experiment.  Spectrometer acceptances were provided by the Hall A Single-Arm Monte Carlo simulation~\cite{Deur,Kramer} used for previous polarized $^{3}$He experiments~\cite{E94010,A1nref}.  Corrections were applied for beam and target polarizations, $P_b$ and $P_t$, and for radiative effects using a modified version of the code POLRAD 2.0~\cite{POLRAD} for internal corrections, and the formalism in Refs.~\cite{MoTsai, Stein} for external corrections.  Input to the radiative corrections came from fits to measured polarized cross-sections  in the quasi-elastic region~\cite{Slifer}, fits to measured $g_1^n$ and $g_2^n$ data in the resonance region~\cite{E94010} and a fit to $g_1^n/F_1^n$~\cite{A1nref}, along with calculations of $g_2^{WW}$, in the deep-inelastic region.  Our measured data were used to guide the fits near $x\simeq0.2$, and the systematic uncertainty in the radiative corrections is dominated by uncertainties in the fits where data are sparse.   

To obtain neutron spin structure functions, a correction was applied to the measured values of $g_1^{^3{\rm{He}}}$ and $g_2^{^3{\rm{He}}}$ based on a model $^3$He wavefunction~\cite{Bissey},
\begin{equation}
g_{1,2}^n=\frac{1}{P_n+0.056}\left[g_{1,2}^{^3\mathrm{He}} + \left(0.014-2P_p\right)g_{1,2}^p\right],
\end{equation}
where  $P_p=-0.028^{+0.036}_{-0.020}$ and $P_n=0.86^{+0.009}_{-0.004}$ are the effective proton and neutron polarizations in $^3$He~\cite{Bissey,  Ciofi, Nogga}.
For calculations of $g_1^p$ we used the average of scenarios 1 and 2 in the Bl{\"{u}}mlein and B{\"{o}}ttcher (BB) NLO fit to world data~\cite{BB}, evolved to our $Q^2$ using the DGLAP procedure~\cite{DGLAP1, DGLAP2, DGLAP3, DGLAP4}.  For $g_2^p$, values for $g_2^{WW}$ were calculated using BB and Eq.~(\ref{eq:g2ww}).
\begin{table}[h]
\caption{Results for $g_1^n$ and $g_2^n$ from this experiment.  The two uncertainties are statistical and systematic, respectively.}
\begin{tabular}{c|c|c|c|c}
\hline
\hline
$Q^2$  & $x$ & $E$  & $g_1^n$ & $g_2^n$\\
$({\rm{GeV}}/c)^2$ &  & (GeV) & $(10^{-3})$ & $(10^{-3})$\\
\hline
0.57 & 0.16 & 3.465 & $-71.9 \pm5.9\pm8.3$ & $78.3 \pm9.0\pm11.0$ \\
0.78 & 0.18 & 4.598 & $-55.7 \pm7.5\pm7.2$ & $75.2 \pm13.0\pm10.1$ \\
0.94 & 0.19 & 4.598 & $-51.2 \pm5.9\pm7.4$ & $67.5 \pm9.0\pm7.2$ \\
1.13 & 0.19 & 5.727 & $-42.6 \pm5.3\pm7.0$ & $48.3 \pm8.9\pm6.0$ \\
1.34 & 0.20 & 5.727 & $-36.9 \pm5.5\pm6.8$ & $55.0 \pm9.7\pm6.8$ \\
\hline
\hline
\end{tabular}
\label{results}
\end{table}

Results for $g_1^n$ and $g_2^n$ from this experiment are given in Table~\ref{results}.  The largest contributions to the systematic uncertainties in $g_1^n$ and $g_2^n$ come from the uncertainties in the cross-section differences, and, for $g_2^n$,  from uncertainties in the models for the transversely polarized quasi-elastic and resonance-region tails used for radiative corrections.

\begin{figure}[h]
\includegraphics[angle=-90,width=3.4in]{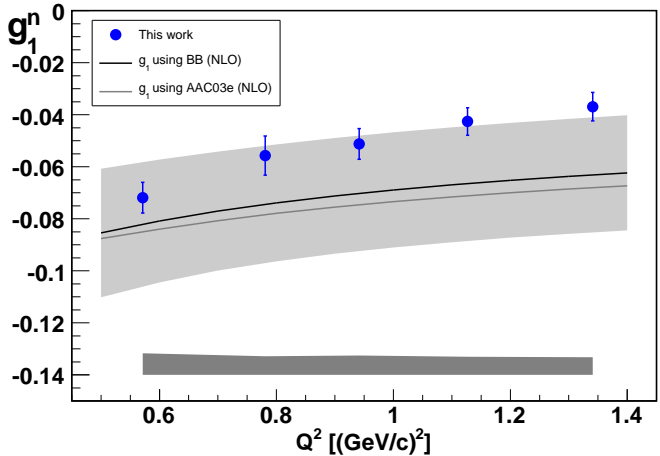}
\caption{Results for $g_1^n$ as a function of $Q^2$.  Data are shown with statistical uncertainties only, with the systematic uncertainty indicated by the lower, dark gray band.  The solid dark line, with gray uncertainty, and the light gray line are calculations using NLO fits to world $g_1^n$ data, evolved to our measured $Q^2$ range.}
\label{g1_data}
\end{figure}

Figure~\ref{g1_data} shows our data for $g_1^n$ as a function of $Q^2$.  Also shown are $g_1^n$ predictions from the NLO analyses of BB and M. Hairi {\em{et al.}}(AAC03e)~\cite{AAC03}.  The uncertainty in the BB curve comes from propagating the uncertainties in the parton distribution functions used in the fit.  The world data for $g_1^n$ at our $x$ are at significantly larger $Q^2$ where higher-twist contributions should be suppressed.  The good agreement with these evolved fits indicates that, within the uncertainties, we are not seeing higher-twist effects in $g_1^n$. This also gives confidence that we can use the BB fit to calculate $g_2^{WW}$ without introducing significant higher-twist effects from $g_1$.  
\vspace*{0.1in}

\begin{figure}[h]
\includegraphics[angle=-90,width=3.4in]{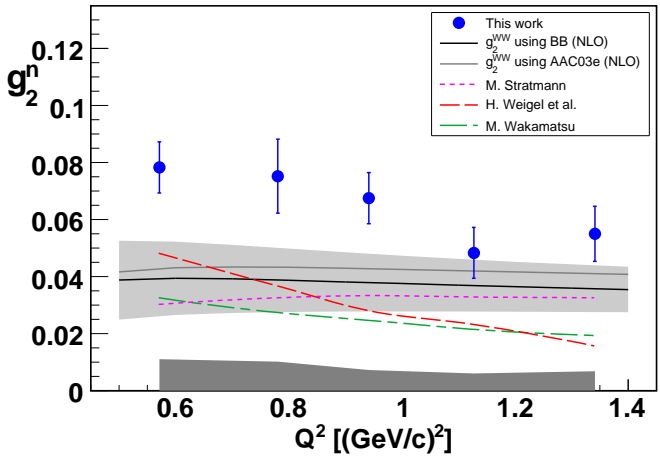}
\caption{Results for $g_2^n$ as a function of $Q^2$ are shown along with model calculations (see text).  Data are shown with statistical uncertainties only, with the systematic uncertainties indicated by the lower, dark gray band. The dark solid line, with gray uncertainty band, and the light gray line are calculations of $g_2^{WW}$ using  NLO fits to world $g_1^n$ data, evolved to our measured $Q^2$.}
\label{g2_data}
\end{figure}

\begin{figure}[h]
\includegraphics[angle=-90,width=3.4in]{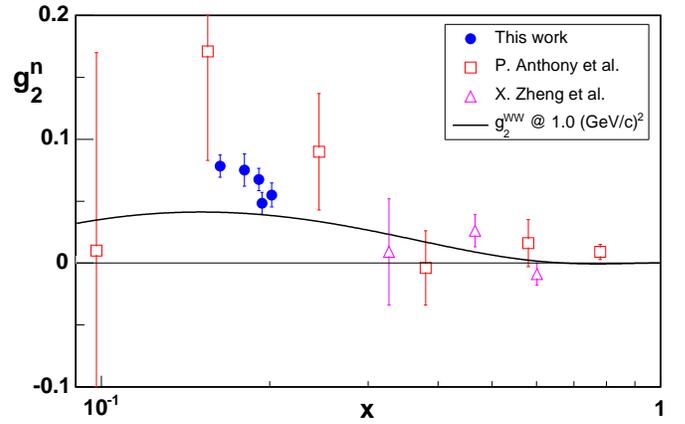}
\caption{Results for $g_2^n$ as a function of $x$ for this experiment are shown along with previous measurements (see text).   Data are shown with statistical uncertainties only.  Results have been slightly displaced in $x$ from the measured values to avoid overlap.  The solid black curve shows $g_2^{WW}$, without uncertainties, at $Q^2=1.0$ $({\rm{GeV}}/c)^2$.}
\label{g2_world}
\end{figure}

Figure~\ref{g2_data} shows our data for $g_2^n$ as a function of $Q^2$ along with calculations of $g_2^{WW}$ from the BB and AAC03e NLO analyses.   The data are more than $5\sigma$ above zero, and at lower $Q^2$, show a systematic positive deviation from $g_2^{WW}$, which we interpret as evidence for non-zero higher-twist contributions.  In the OPE, the twist-2 and twist-3 contributions to $g_2$ enter at the same order in $Q^2$, with additional higher-twist contributions suppressed by powers of $1/Q$.   Assuming these additional higher-twist contributions are small, we expect the quantity $g_2-g_2^{WW}$ to be constant as a function of $Q^2$.  When compared to $g_2^{WW}$ from BB, a fit to our data gives $g_2-g_2^{WW}=0.0262 \pm 0.0043 \pm 0.0080 ±\pm0.0099$, with a reduced $\chi^2$ of 1.4.  The first two uncertainties are from the experiental statistical and systematic uncertainties and the third is from the uncertainty in the BB fit.   Attempts to fit the data with functions that contain additional higher-twist dependence and/or logarithmic perturbative QCD corrections did not improve the quality of the fits.  

Also shown in Figure~\ref{g2_data} are chiral soliton model calculations from Weigel {\em{et al.}}~\cite{Weigel1, Weigel2} and Wakamatsu~\cite{Wakamatsu}, and a bag model calculation of the higher-twist contribution from Stratmann~\cite{Strat}, combined with  $g_2^{WW}$ from BB.  Our data indicate a positive contribution from higher-twist effects while the model calculations generally predict a negative contribution.

Figure~\ref{g2_world} shows the improved quality of our  $g_2^n$ data (solid points) plotted versus $x$, as compared to older data (open squares) from the Stanford Linear Accelerator Center~\cite{E155x}.  Also shown are recent data (open triangles) from Jefferson Lab  at higher $x$~\cite{A1nref}.  The data shown cover a range from  $Q^2=0.7$ $({\rm{GeV}}/c)^2$ at the lowest $x$ to $Q^2=20$ $({\rm{GeV}}/c)^2$ at the highest $x$.    For reference, $g_2^{WW}$ from BB is also included at constant $Q^2=1.0$ $({\rm{GeV}}/c)^2$.

In summary, we have made the first measurement of the $Q^2$-dependence of the $g_2$ spin structure function for the neutron at $x\simeq0.2$ in the range $Q^2=0.57-1.34$ $({\rm{GeV}}/c)^2$.   With a factor of $>10$ improvement in statistical precision over previous measurements, these data allow for an accurate determination of the higher-twist contributions by direct comparison with the twist-2 $g_2^{WW}$ prediction.  Our results show a positive deviation from $g_2^{WW}$ at lower-$Q^2$, indicating a non-zero higher-twist contribution, and are inconsistent with model calculations, which generally predict a negative higher-twist contribution.   Precision data for $g_1$  obtained in this kinematic range showed good agreement with the NLO analyses of world data, indicating no significant higher-twist contributions within the uncertainties. 

\acknowledgments{
We  thank the personnel of Jefferson Lab for their support and efforts towards the successful completion of this experiment.  We would also like to thank M. Stratmann, H. Weigel, L. Gamberg and M. Wakamatsu for additional model calculations at our kinematics.  This work was supported by the Department of Energy (DOE), the National Science Foundation, the Jeffress Memorial Trust, the Italian Istituto Nazionale di Fisica Nucleare, the French Institut National de Physique Nucl\'{e}aire et de Physique des Particules/CNRS and the French Commissariat \`{a} l'Energie Atomique.   The Southeastern Universities Research Association operates the Thomas Jefferson National Accelerator Facility for the DOE under contract No. DE-AC05-84ER40150.
}

\end{document}